\documentclass{aastex}
\usepackage{emulateapj5,times,mathptm}
\singlespace
\slugcomment{Accepted for the publication in the {\it Astrophysical Journal Letters}}
\def\la{\mathrel{\hbox{\rlap{\hbox{\lower4pt\hbox{$\sim$}}}\hbox{$<$}}}}
\def\ga{\mathrel{\hbox{\rlap{\hbox{\lower4pt\hbox{$\sim$}}}\hbox{$>$}}}}

\shortauthors{Park}
\shorttitle{G292.0+1.8}

\begin{document}
\title{The Structure of the Oxygen-rich Supernova Remnant G292.0+1.8 
from {\it Chandra} X-ray Images: Shocked Ejecta and Circumstellar Medium}
\author{Sangwook Park\altaffilmark{1}, Peter W. A. Roming\altaffilmark{1}, 
John P. Hughes\altaffilmark{2}, Patrick O. Slane\altaffilmark{3}, 
David N. Burrows\altaffilmark{1}, Gordon P. Garmire\altaffilmark{1}, and
John A. Nousek\altaffilmark{1} }

\altaffiltext{1}{Department of Astronomy and Astrophysics, Pennsylvania State
University, 525 Davey Laboratory, University Park, PA. 16802; 
park@astro.psu.edu}
\altaffiltext{2}{Department of Physics and Astronomy, Rutgers University,
136 Frelinghuysen Road, Piscataway, NJ. 08854-8109}
\altaffiltext{3}{Harvard-Smithsonian Center for Astrophysics, 60 Garden Street,
Cambridge, MA. 02138}

\begin{abstract}

We present results from the observation of the young Galactic
supernova remnant (SNR) G292.0+1.8 with the Advanced CCD Imaging
Spectrometer (ACIS) on board the {\it Chandra X-ray Observatory}.  In
the 0.3 $-$ 8 keV band, the high resolution ACIS images reveal a
complex morphology consisting of knots and filaments, as well as the
blast wave around the periphery of the SNR. We present
equivalent width (EW) maps for the elemental species O, Ne, Mg, and
Si, which allow us to identify regions of enhanced metallicity in the
SNR. G292.0+1.8 is bright in O, Ne, and Si; weaker in S and Ar;
with little Fe. The EW and broad-band images
indicate that the metal-rich ejecta are distributed primarily
around the periphery of the SNR. The central belt-like 
structure has normal solar-type composition, strongly suggesting 
that it is primarily emission from shocked circumstellar medium 
rather than metal-rich ejecta. We propose that the belt traces its 
origin to enhanced mass loss in the star's equatorial plane during 
the slow, red supergiant phase. We also identify thin filaments 
with normal composition, centered on and extending nearly 
continuously around the outer boundary of the SNR. These may 
originate in a shell caused by the stellar winds from the
massive progenitor in the red/blue supergiant phases, over-run
by the blast wave. 
\end{abstract}


\section {\label {sec:intro} INTRODUCTION}

G292.0+1.8 is a bright Galactic SNR, discovered in the radio 
surveys of the southern sky \citep{milne69,shaver70}. 
The radio morphology is centrally-peaked with an extended 
fainter plateau \citep{lockhart77,braun86} and the radio spectral
index is intermediate between shell and plerion type
SNRs \citep{goss79}. The detection of strong O and Ne lines
in the optical spectrum \citep{goss79,murdin79}
classified G292.0+1.8 as an O-rich SNR, of which there are only 
two other examples in the Galaxy: Cassiopeia A and Puppis A. 
G292.0+1.8 is a young SNR ($\la$ 1600 yr) \citep{murdin79}, 
making it useful for studying the processes by which SN ejecta 
mix with the ambient ISM.

G292.0+1.8 was first detected in X-rays with {\it HEAO}-1 
\citep{share78} and a central bar-like structure surrounded by 
an ellipsoidal disk was observed with {\it Einstein} \citep{tuohy82}.
X-ray spectra taken with {\it Einstein} and {\it EXOSAT} 
revealed strong atomic emission lines from highly ionized Mg, Si, 
and S \citep{clark80,claas88}. Hughes and Singh (1994) performed 
a nonequilibrium ionization (NEI) analysis of the X-ray spectrum of
G292.0+1.8, which determined that the SNR was significantly enhanced 
in O, Ne, and Mg and that the pattern of elemental abundances agreed 
well with the nucleosynthesis expected from a massive star (20$-$25 
$M_\odot$) core collapse supernova.

The superb high angular resolution of the {\it Chandra X-ray
Observatory} \citep{weisskopf96} allows us to study 
the complex structure of the ejecta and the surrounding medium.
It also allowed us to discover a hard point-like 
source near the center of the remnant, as reported elsewhere
\citep{hughes01a} (H01a, hereafter). Here we report the first 
{\it Chandra} imaging results on the diffuse thermal emission 
from G292.0+1.8. A subsequent {\it Letter} describes our 
spectral analysis of the SNR \citep{hughes01b} (H01b, hereafter).

The observation is described in \S\ref{sec:obs}. 
Data analysis and results are presented in \S\ref{sec:images} 
and the implications are discussed in \S\ref{sec:disc}.
A summary is presented in \S\ref{sec:summary}.

\section{\label{sec:obs} OBSERVATION \& DATA REDUCTION}

G292.0+1.8 was observed with the Advanced CCD Imaging Spectrometer 
(ACIS) on board {\it Chandra} on 2000 March 11 as part of the 
{\it Chandra} Guaranteed Time Observation program. The angular size 
of G292.0+1.8 is slightly larger than the 8$'$ size of a single CCD 
chip; the ACIS-S3 chip was chosen because it has the best energy 
resolution over the entire CCD. The pointing was selected to place 
most of the SNR on the S3 chip.

We have utilized new data reduction techniques developed
at Penn State for correcting the spatial and spectral degradation
of ACIS data caused by radiation damage, known as Charge Transfer
Inefficiency (CTI) \citep{townsley00}. 
The expected effects of the CTI correction include an increase in
the number of detected events and improved event energies and 
energy resolution \citep{townsley00,townsley01}.
We screened the data with the flight timeline filter and then
applied the CTI correction before further data screenings by
status, grade, and the energy selections. ``Flaring'' pixels were
removed and {\it ASCA} grades (02346) were selected. Events with
inferred energies between 0.3 keV and 8.0 keV were extracted for 
further data analysis. The overall lightcurve was examined 
for possible contamination from time variable 
background and no significant variability was found.
After data processing $\sim$43~ks of effective exposure was left.

\section{\label{sec:images} X-ray Images \& Equivalent Width Maps}

\begin{figure}[]
\figurenum{1}
\centerline{\includegraphics[angle=0,width=7cm]{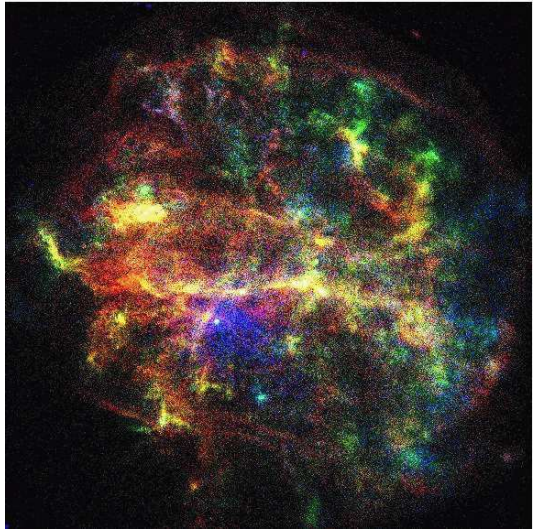}}
\figcaption[]{The ``true-color'' image of 
G292.0+1.8 from the {\it Chandra}/ACIS observation. Red, green, 
and blue represent 0.3 $-$ 0.8 keV, 0.85 $-$ 1.7 keV, and 1.7 
$-$ 8.0 keV, respectively. Red and blue are weighted the same 
while green is weighted 5.3 times the other two colors. 
The color palette is the weighted sum of these three colors.
A linear scaling was used. This image is from the entire 
ACIS-S3 chip and parts of the remnant extend beyond the chip boundary. 
\label{fig:fig1}}
\end{figure}

Figure~\ref{fig:fig1} shows an X-ray ``true-color" image of 
G292.0+1.8, with red indicating the lowest energies and 
blue for the highest. The deep blue region south of the 
central belt is the presumed pulsar wind nebula, 
dominating above 2 keV (H01a). Below 2 keV, the SNR is 
dominated by a complex network of knots and filaments (yellow 
and green) on angular scales down to the instrumental 
resolution. The striking color contrasts in this filamentary 
structure indicate spectral variations in this material, 
produced by varying line strengths. These knots and filaments 
are bounded by shell-like circumferential filaments 
surrounding most of the SNR, outside of which is faint 
diffuse emission. These features are soft, as indicated 
by the red colors.

The overall ACIS spectrum (H01a) shows broad 
atomic emission line complexes from the elemental species O, 
Ne, Mg, Si, S, and Ar, while little emission from Fe is 
observed. In order to trace the elemental distribution
across the SNR, we constructed {\it equivalent 
width} (EW) images by selecting photons around the broad line 
complexes (Table \ref{tbl:tab1}), using a technique pioneered
by Hwang et al. (2000). Line and continuum bandpasses were 
selected for each spectral line of interest.
Images in these bands were extracted with 2$\arcsec$ pixels 
and smoothed with a Gaussian with $\sigma$ = 7$\arcsec$.
The underlying continuum was calculated by logarithmically 
interpolating between images made from the higher and lower 
energy ``shoulders'' of each broad line.
The estimated continuum flux was integrated 
over the selected line width and subtracted from the line 
emission. The continuum-subtracted line 
intensity was then divided by the estimated continuum
on a pixel-by-pixel basis to generate the EW images for each 
element. In order to avoid noise in the EW maps caused by
poor photon statistics near the edge of the remnant,
we have set the EW values to zero where the estimated
continuum flux is low. We also set the EW to zero where
the integrated continuum flux is greater than the line flux.

\begin{figure}[]
\figurenum{2}
\centerline{\includegraphics[angle=-90,width=14cm]{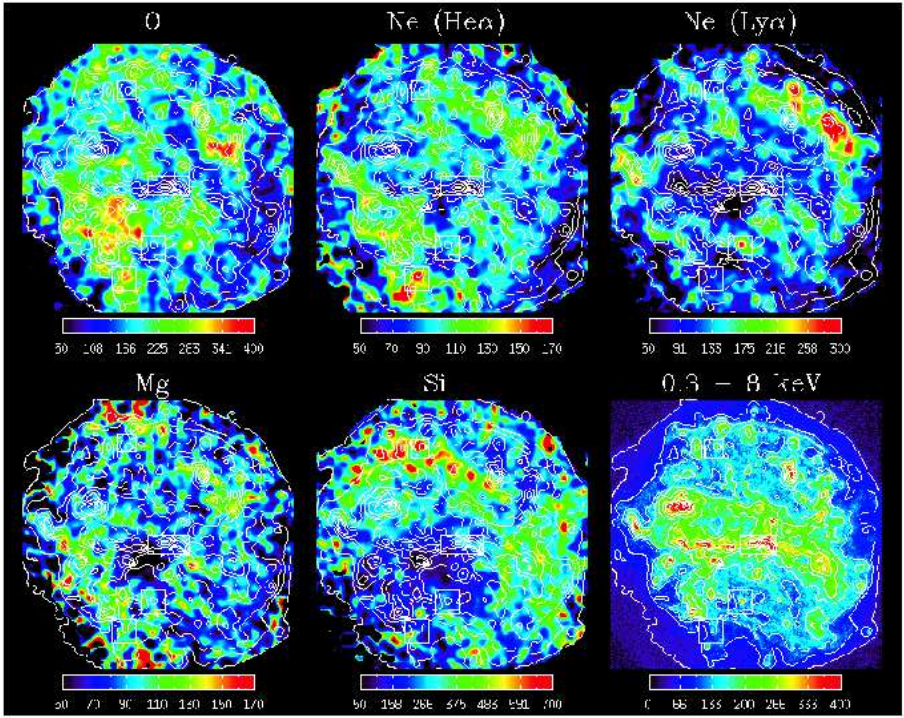}}
\figcaption[]{The equivalent width (EW) images for 
the elemental species O, Ne, Mg, and Si. The false-color
scales are in units of eV for the EW maps and are 
counts pixel$^{-1}$ for the broad band image. The line and continuum 
images have been smoothed by convolving with a $\sigma$ = 7$\arcsec$ 
Gaussian prior to calculation of the EW. Contours of the
0.3 $-$ 8 keV band surface brightness are overlaid in each
image. Zero EW has been assigned where the estimated underlying
continuum is $\leq$10\% of the mean. For comparison with Figure 3b, 
locations for regions of sample spectra are marked with rectangles. 
Actual extraction regions are presented in Figure 3b.
\label{fig:fig2}}
\end{figure}

\begin{figure}[]
\figurenum{3}
\centerline{{\includegraphics[angle=-90,width=7cm]{fig3a.ps}}
{\includegraphics[angle=-90,width=8.5cm]{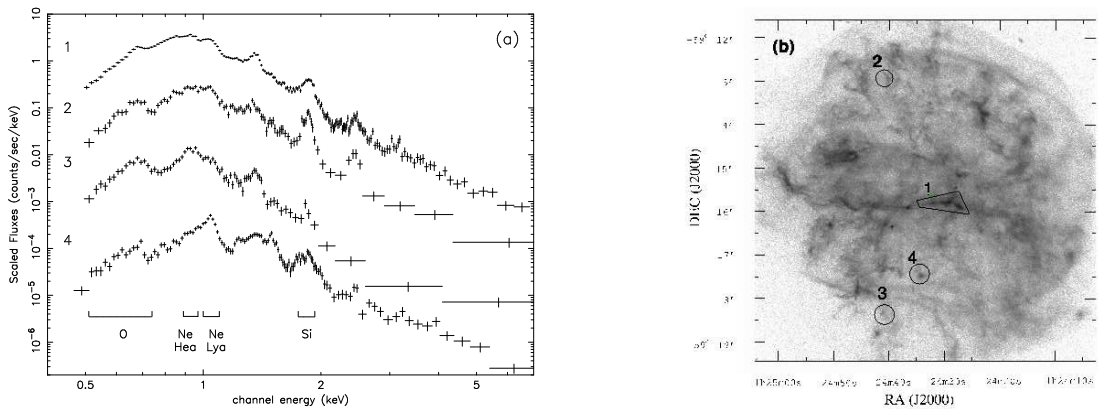}}}
\figcaption[]{(a) Spectra from selected regions within G292.0+1.8.
The fluxes are arbitrarily scaled to allow comparisons of
the overall spectral features among the regions. The 
energy bands used to produce the EW maps are indicated.
(b) The selected regions for which spectra are displayed in panel (a).
The gray-scale image is the 0.3 $-$ 8 keV surface brightness image.
The selected regions were chosen for:
1: bright broad-band emission from the equatorial belt,
2: high Si EW,
3: high Ne He$\alpha$ EW, and
4: high He Ly$\alpha$ EW.
\label{fig:fig3}}
\end{figure}

The EW maps for O, Ne, Mg, and Si are presented in Figure
\ref{fig:fig2} with overlaid contours from the 0.3 $-$ 8.0 keV
band image for morphological comparison between the elemental
distribution and the surface brightness.
(The S and Ar EW maps are too noisy to be useful. 
High Mg EW values in the extreme north and south edges of 
the SNR are due to low continuum and are insignificant.)
The overall spectrum exhibits two broad-line features for Ne
(He$\alpha$ at $\sim$0.93 keV and Ly$\alpha$ at $\sim$1.05 keV)
(H01a), and we separately present those two line components
in the EW maps. The morphology of the EW maps depends on
the species, and does not necessarily trace the overall 
broad-band morphology. To investigate the possibility that 
complex broad line features might have resulted in contamination 
of the estimated underlying continuum, we have compared the O EW 
map with the optical O{\small II} $\lambda$ 3727$\AA$ line 
data \citep{tuohy82}. 
The O EW enhancements in the southeast area of the SNR are 
in good spatial agreements with the optical data, supporting 
the reliability of the EW maps. We also extracted a few 
sample spectra from small regions across the SNR in order
to make a more realistic test for the overall validity
of the metallicity structure of the SNR implied by the EW maps.
The metallicity distributions implied by the EW maps are in 
good agreement with the actual spectra (Figure \ref{fig:fig3}). 
For example, the spectrum in Region 1 shows weaker Ne lines compared
with those from Regions 3 and 4, in agreement with the EW maps.
The spectrum of Region 2, which sits on a peak in the Si EW map, 
is dominated by bright Si and S lines.
We thus conclude that our EW maps are 
reliable despite the inherent uncertainties in the continuum 
subtraction. Further quantitative support for the
validity of our EW maps is given in H01b.

The EW maps reveal a variety of chemical structures and 
ionization states within the SNR, which vary with the 
species. The O EW map indicates strong O emission throughout 
the SNR, particularly in the southeastern quadrant, consistent 
with the previously-known O-rich nature. The Ne EW maps are 
particularly interesting. Both Ne maps are enhanced within 
the circumferential filaments, suggesting that the material 
interior to these filaments is dominated by ejecta. The bright, 
broad ridge of emission in the northwest corner of the Ne 
Ly$\alpha$ map lies to the outside of the corresponding 
feature in the Ne He$\alpha$, indicating a progression 
in ionization state with distance from the shock boundary 
in this portion of the SNR. This may be attributed to the
progressive ionization by the reverse shock as seen 
in SNR 1E0102.2$-$7219 \citep{gaetz00,flanagan01}.
In the southeastern quadrant, O and Ne He$\alpha$ are both 
strong, while Si and Ne Ly$\alpha$ are relatively weak, again 
suggesting large-scale variations in ionization state and 
elemental abundance within the ejecta. The central bar that 
is prominent in the broad-band images is either a trough 
or completely absent in the EW maps, indicating that 
this feature is not related to the ejecta, but is a 
structure in the shocked circumstellar material (CSM). The 
thin circumferential filaments are also absent from the EW 
maps, indicating a CSM origin. 
Finally, the region of the pulsar wind nebula surrounding the 
bright point-like source is dominated by synchrotron emission 
(H01a), and is therefore dark in the EW maps.

\section{\label{sec:disc} DISCUSSION}

The most interesting result from the EW and the surface
brightness maps is the inferred CSM origin of the bright 
belt-like structure as observed in the broad-band image. 
Tuohy et al. (1982) suggested that the central ``bar-like'' 
enhancement results from the equatorial expulsion of 
O-rich ejecta from a rotating massive progenitor to form 
an expanding ring of such ejecta. Our results from this
{\it Chandra} observation indicate that instead of being 
O-rich, the material associated with the belt is actually 
O-poor compared with the rest of the remnant.
This {\it equatorial belt} is most likely
emission from the normal composition material into which the 
blast wave is expanding, as suggested by the EW maps, 
and is strong evidence for the existence 
of a non-spherically symmetric CSM produced by the stellar winds 
from the massive progenitor (Blondin et al. 1996 and references
therein). Such asymmetric wind structures have been invoked to 
explain complex biannular structure in some radio SNRs 
\citep{manchester87}, including G296.8-00.3 \citep{gaensler98}.
The apparent belt-like morphology for G292.0+1.8 
suggests that the presumed asymmetric CSM may be ``ring-like'', 
reminiscent of the well-known optical ring of SN 1987A 
\citep{luo91}, but seen edge-on and on much larger spatial scales. 
Assuming that the apparent ``length'' of the equatorial 
belt represents the diameter of such a circumstellar ring,
the current size of the belt is $\sim$3 pc in radius at the 
distance of 4.8 kpc. This is comparable with the observed sizes 
of the circumstellar nebulae from massive stars ($\la$   
a few pc in radius depending on their masses and the 
evolution history; e.g., Chu 2001 and references therein). 

Another interesting feature is the soft circumferential filaments 
that nearly enclose the ejecta-dominated portion of the SNR.  
Although the material just interior to these filaments shows 
strong lines of O, Ne, and Si, the circumferential filaments 
themselves are absent from the EW maps, indicating low line 
strength. This is consistent with an origin in the CSM, but 
several scenarios may be invoked to explain this large-scale 
feature. The low abundances and thin shell suggest the 
possibility that these filaments represent a relic structure 
in the CSM, perhaps a shell at the boundary of a red supergiant 
wind with a subsequent blue supergiant wind, that has been 
over-run by the blast wave. Numerical simulations of such 
CSM/blast wave interactions (e.g., Franco et al. 1991) have
shown that the observed structure can be produced if the mass
of the shell is low. Alternatively, it could represent the 
outer shock itself. This interpretation would require that the 
low surface-brightness emission exterior to these filaments, 
results from projection effects, which could happen if 
one hemisphere of the remnant was expanding into a lower density 
medium, producing a morphology similar to VRO 42.05.01 
\citep{landecker82,pineault87,burrows94}, but seen face-on. 

We note that H01a has attributed the $\sim$1$'$ offset
of the pulsar position from the geometrical center of the SNR
to either high transverse velocity of the pulsar or slower
expansion of the SNR toward the southeast. Our results
suggest advanced ionization state in the northwest and 
thus likely a higher ISM density there, which would support 
the high pulsar velocity interpretation. 

\section{\label{sec:summary} SUMMARY}

We have presented results from the {\it Chandra}/ACIS 
observation of the Galactic O-rich SNR, G292.0+1.8. 
The high resolution ACIS images resolve the complex 
filamentary structure in the soft band while revealing 
a pulsar and its wind nebula in the hard band. The 
EW maps 
indicate variable elemental abundance and ionization 
across the SNR. Based on the surface brightness images
and the EW maps, we propose that 
the equatorial belt in the broad band images is emission 
from the shocked non-spherically symmetric CSM,
produced by slow stellar winds from the massive 
progenitor in the red supergiant phase, rather than from 
metal-rich ejecta. We also suggest that the origin of the 
soft shell-like structure around the outer boundary of 
the SNR may be emission from the shell of the stellar 
winds over-run by the blast wave, or the shock front itself 
if the X-ray emission of larger radii comes from portions 
of the remnant seen in projection along the line of sight.
The distribution of the ejecta implies a complex evolution
for G292.0+1.8, such as an asymmetric SN explosion and
highly clumped ejecta interacting with non-uniform ambient
ISM \citep{braun86}.

We resolve the complex X-ray structure of G292.0+1.8: 
i.e., metal-rich ejecta, shocked dense asymmetric CSM, 
and the shocked stellar winds or ISM
with the high resolution ACIS data.
We should note, however, that EW depends not only on the
abundance, but also on the temperature and ionization state,
so that direct spectral analysis is needed to confirm the
inferred metallicity distributions by the EW maps.
Further progress on the origin of the equatorial belt
and the soft shell-like structure at the SNR's periphery
will benefit from follow-up optical spectroscopy
and detailed X-ray spectral analysis (H01b).

\acknowledgments
{The authors thank L. Townsley and colleagues in the department of
Astronomy \& Astrophysics at Penn State University for developing
the software for the CTI correction of the ACIS data.
We also thank K. Lewis for her help at the early stage of this work.
This work was funded by NASA under contract NAS8-38252 to Penn State,
NAS8-39073 to CfA and {\it Chandra} grant GO0-1035X to Rutgers University.
}

\clearpage

\begin{deluxetable}{cccc}
\footnotesize
\tablecaption{Energy Bands used for Generating the Equivalent Width
Images. 
\label{tbl:tab1}}
\tablewidth{0pt}
\tablehead{ \colhead{Elements} & \colhead{Line} & 
\colhead{Low\tablenotemark{\dag}} & \colhead{High\tablenotemark{\dag}} \\
 & \colhead{(eV)} & \colhead{(eV)} & \colhead{(eV)} }
\startdata
O & 510 $-$ 740 & 300 $-$ 510 & 740 $-$ 870 \\
Ne He$\alpha$ & 890 $-$ 970 & 740 $-$ 870 & 1120 $-$ 1160 \\
Ne Ly$\alpha$ & 1000 $-$ 1100 & 740 $-$ 870 & 1120 $-$ 1160 \\
Mg & 1290 $-$ 1420 & 1250 $-$ 1290 & 1620 $-$ 1700 \\
Si & 1750 $-$ 1930 & 1620 $-$ 1700 & 2020 $-$ 2120 \\
\enddata
\tablenotetext{\dag}{The high and low energy bands around the
selected line energies used to estimate the underlying continuum.}

\end{deluxetable}


\end{document}